\documentclass[12pt]{iopart}
\usepackage{iopams}  
\usepackage{graphicx}

\begin{document}
\title{Spin-spin correlations of entangled qubit pairs in the Bohm interpretation of quantum mechanics}

\author{A. Ram\v{s}ak}
\address{Faculty of Mathematics and Physics, University of  
Ljubljana, Ljubljana, Slovenia}
\address{J. Stefan Institute, Ljubljana, Slovenia}
\ead{anton.ramsak@fmf.uni-lj.si}
\begin{abstract}
A general entangled qubit pair is analyzed in the de Broglie-Bohm formalism corresponding to two spin-1/2 quantum rotors. 
Several spin-spin correlators of Bohm's hidden variables are analyzed numerically and a detailed comparison with results obtained by standard quantum mechanics is outlined. In addition to various expectation values the Bohm interpretation allows also a study of the corresponding probability distributions, which enables a novel understanding of entangled qubit dynamics. In particular, it is shown how the angular momenta of two qubits in this formalism can be viewed geometrically and characterized by their relative angles. For perfectly entangled pairs, for example, a compelling picture is given, where the qubits exhibit a unison precession making a constant angle between their angular momenta.
It is also  demonstrated that the properties of standard quantum mechanical  spin-spin correlators responsible for the violation of Bell's inequalities  are identical to their counterparts emerging from the probability distributions obtained by the Bohmian approach.    
\end{abstract}

% keywords here, in the form: keyword \sep keyword
%Bohm hidden variables \sep causal theory of spin-1/2 \sep  Bell's inequalities

%\pacs{03.65.Ca, 03.65.Ud, 03.67.Bg}
%Quantum systems with finite Hilbert space
%Entanglement and quantum nonlocality
%Entanglement production and manipulation

\maketitle
\section{Introduction}

It is well known that several phenomena pertinent to spin-$\frac{1}{2}$ systems can not be described in terms of classical variables although to some extent a geometric terminology may be applied to a spin state of a single electron which is customary presented on the Bloch sphere, where the polar and azimuthal angles can be viewed as the Euler angles of a unit vector pointing along the spin direction. Still,  within the framework of standard quantum mechanics there is no answer to the questions like: How could one imagine   spin properties of an entangled electron pair in terms of classical variables? The question is related to the problem of quantum entanglement, which is a rather mysterious concept and has remained so since it was first introduced by Schr{\" o}dinger back in 1935. Despite this, with the recent surge in interest stimulated by quantum information processing \cite{compute,vedral} and quantum sensing below the quantum limit \cite{limit}, discussions about the meaning of entanglement and how it might be visualized beyond the standard quantum mechanics have once more come to the fore \cite{einstein,valentini}.  In this context can also be set recent intense exchange of views  whether the wavefunction is physically real and can be directly measured  or represents only a statistical tool that reflects our ignorance of the particles being measured \cite{wave}. 

Quantum entanglement is strongly related to the discussion of locality in quantum mechanics and with the introduction of Bell's inequality also to the question of the existence of hidden variables  \cite{bell64,bell87}. In order to set  experimentally observable boundaries to various hidden variable theories several other inequalities have been introduced \cite{gisin03,gisin07}. The outcome of numerous experimental tests  of such inequalities in the last three decades performed on two-particle systems ruled out all local hidden variable approaches, although it seems that non-locality concept itself is not sufficient to be consistent with quantum experiments \cite{bloblacher07}. However, for multipartite entangled states local hidden variable models can be introduced such that the violation of most of the Bell-type inequalities does not imply and new inequalities have to be established \cite{toth06,bancal11}. 

In view of the fact that quantum entanglement is not an observable according to the usual rules of quantum mechanics, it does not have a direct classical analogue. However, some insight can be given by a description based on the pilot-wave approach of de Broglie \cite{debroglie} and  later developed by Bohm \cite{bohm1} -- where an effective non-local potential is introduced along with hidden variables. This theory is manifestly non-local, {\it i.e.}, non-local potentials 
represent non-local action which does not get small with spatial distance between the particles and this is
the key ingredient which guarantees predictions in accordance with standard quantum mechanics. 
While it seems that such an introduction of non-local variables does not lead to new predictions -- it might turn out to be interesting in the context of possible device independent analysis of entangled systems \cite{pironio,bancal11}. In particular, one can ask questions related to pre-existing properties of the system; explicitly, what each individual particle in the observed system is  doing prior performing the measurement. 

Since the introduction of the Bohm interpretation of quantum mechanics
it has been in detail explored how this approach leads to outcome predictions of various experiments identical to the results of standard quantum mechanics, and gives due to its nature, additionally also the interpretation of the motion of particles in terms of particular  coordinates -- commonly termed as the Bohm hidden variables. After the establishment, the Bohm formalism was extended from original single spinless particles to many-body fermionic or bosonic systems  \cite{holland00,durr,riggs} and also to the quantum field theory, including
creation and annihilation of particles \cite{dur04,nikolic}. Recently it was shown
how the Born-rule probability densities of non-relativistic quantum mechanics emerge naturally from the particle dynamics of de Broglie-Bohm  
theory \cite{towler}. The interpretation of several typically quantum experiments was developed,
for example, for the double slit experiments or tunneling of particles through barriers \cite{holland00}, the Aharonov-Bohm effect \cite{abeffect}, and Stern-Gerlach experiments \cite{stern}. The Bohmian theory in
terms of well-defined individual particle trajectories with continuously variable spin vectors
was successfully applied also to the problem of  Einstein-Podolsky-Rosen spin correlations \cite{dewdney}. Among the simplest quantum objects are two state systems and the formal Bohm approach is consistently extended to spin-$\frac{1}{2}$ systems with non-relativistic formalism based on the Pauli equation \cite{pauli}, causal rigid rotor theory \cite{holland88}, via spinor wave functions \cite{durr}, 
the Bohm-Dirac model for entangled electrons \cite{dur99}, by Clifford algebra approach to Schr{\" o}dinger \cite{hiley1} or relativistic Dirac particles \cite{hiley2}.

In this paper we concentrate on the question of the visualization of the angular momenta -- spins -- of a system of two entangled  particles by the  approach of the causal interpretation of a system of two spin-$\frac{1}{2}$  particles as introduced by 
Holland \cite{holland88,holland00} where the starting point is the mapping between the quantum rigid rotor and a spinning top in the presence of a quantum potential. In particular, we apply this ontological formalism to the case of a two-particle state
\begin{equation}
\left| \Psi\right\rangle=\cos { \vartheta \over 2} \left|\uparrow\downarrow\right\rangle+e^{i \varphi}\sin {\vartheta \over 2} \left|\downarrow\uparrow\right\rangle,
\label{psi2}
\end{equation}
where for the sake of convenience we use spin-$\frac{1}{2}$  notation where $\left|\uparrow\downarrow\right\rangle$ corresponds to the state of the system when the first particle (qubit) is in the "up" state, {\it i.e.}, in the direction of the $z$-axis and the second qubit is in the state "down". The generalization to states spanned by $\left|\uparrow\uparrow\right\rangle$ and  $\left|\downarrow\downarrow\right\rangle$ or to systems represented as mixed states is possible, but it is not considered here.
Qubits are not restricted to real spin of electrons and may be realized by any appropriate two state quantum system, as are for example  flux qubits in superconducting rings \cite{makhlin}, charge pseudo-spin of electron pairs in double quantum dots \cite{mravlje}, two flying qubits in quantum point contacts \cite{rejec}, entangled photon pairs \cite{fotoni},  or even most recently studied two-qubit composite systems \cite{delfot}.

The causal approach considered here has been introduced more than two decades ago but exact solutions are known only for the case of a single qubit and for some properties of a qubit pair in the singlet state. It is also known that the Bell inequalities are violated in this formalism and recently the  quantum entanglement of a qubit pair was linked to a particular motion of angular momenta in the Bohmian space of hidden variables in a way which is
entirely missed in the usual quantum mechanical approach \cite{epl11}. The problem of the motion of two entangled rotors in the Bohmian space of hidden variables can be efficiently treated numerically,  but the author is not aware of any quantitative investigation of this problem. The aim of the present paper is to fill the gap by a comprehensive  review of  spin-spin correlations and the corresponding probability distributions of Bohmian hidden variables in a system of two entangled qubits in the quantum state Eq.~(\ref{psi2}). We explore also the relation to the Bell's inequalities for the case of a perfectly entangled qubit pair.

The paper is organized as follows:
after the introduction in Section 2 the model for two quantum rigid rotors in the causal picture is introduced.  In Section 3  numerically extracted probability distributions of various hidden variables related to two spin-$\frac{1}{2}$  rotors and the corresponding ensemble average values are presented. The Shannon entropy for a discretized probability distribution of the angular momentum $z$-axis projection is given. Section 4 is devoted to geometrical aspects of two quantum rotors, in particular, to the study of the relative angles between the angular momenta and  the corresponding fluctuations. In Section 5 we express the Bell's inequalities in terms of the correlators derived in Section 3. Section 6 is devoted to summary and conclusions.

%\begin{figure}[htbp]
\begin{figure}[b]
\begin{center}
\includegraphics[width=120mm]{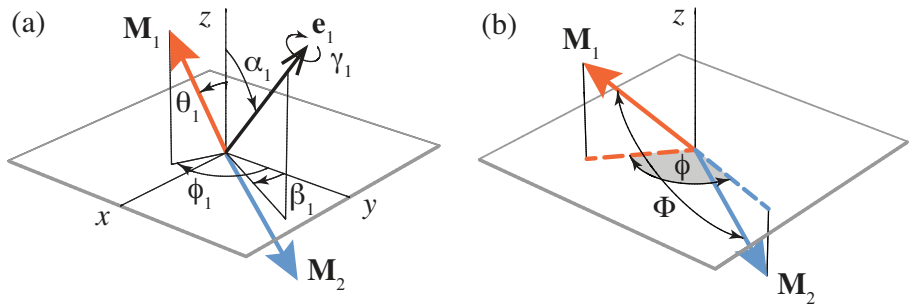}
\caption{(a) Rotor 1 principal axis ${\mathbf{e}}_1$ with Euler angles $\lambda_1=\{\alpha_1,\beta_1,\gamma_1\}$ (${\mathbf{e}}_2$ for rotor 2 is not shown) and  angular momenta ${\mathbf{M}_{1,2}}$, for a general case of two qubits. In the case of unentangled rotors, $\vartheta=0$,  the vectors ${\mathbf{M}_1}$, ${\mathbf{e}}_1$ and the $z$-axis are coplanar, Eq.~(\ref{mzeta}). Euler angles $\theta_{1}$,  $\phi_{1}$  correspond to the angular momentum ${\mathbf{M}}_1$ and  $\theta_{2}$, $\phi_{2}$ (not shown) to ${\mathbf{M}}_2$. (b) Two qubit case 
with shown relative angle $\Phi$ and the $xy$-in-plane azimuthal angle $\phi$ between ${\mathbf{M}}_1$ and ${\mathbf{M}}_2$. }
\label{fig1}
\end{center}
\end{figure}

\section{Bohmian treatment of rigid rotors}

\subsection{A single rotor}

As a preliminary to the detailed examination of entangled rotor pairs we consider 
a single quantum spherically symmetric rigid rotor treated by the Hamiltonian 
\begin{equation}
{\hat{H}}={\hat{\mathbf{M}}^2 \over 2 I},
\label{qH}
\end{equation}
where   $\hat{\mathbf{M}}$  represents the angular momentum operator -- a differential operator whose components are the infinitesimal generators of the rotation group $SO(3)$ and
$I$ is the moment of inertia \cite{holland00,holland88}. Eigenstates $\psi$  of the three mutually commuting operators
${\hat{\mathbf{M}}}^2$, ${\hat{{M_z}}}$ and ${\mathbf{e}}\cdot\hat{\mathbf{M}}$
are functions of  Euler angles $\lambda=\{\alpha,\beta,\gamma\}$, specifying the orientation of a rigid body with the principal axis defined by a normalized vector ${\mathbf{e}}$. The angular momentum eigenvalues are defined by
\begin{eqnarray}
{\hat{\mathbf{M}}}^2\psi&=&s(s+1)\psi,\\
{\hat{{M}}}_z\psi&=&m\psi,\\
{\mathbf{e}}\cdot\hat{\mathbf{M}}\psi&=&n\psi,\label{edotm}
\end{eqnarray}
where $s$ is the total angular momentum quantum number, and $m$, $n$ are the angular momentum projection quantum numbers: in the spatial $z$-axis direction and with respect to the rotor axis ${\mathbf{e}}$, respectively.

In order to analyze ontological properties of rigid rotors in de Broglie-Bohm space of hidden variables, we in this Section closely
follow  the formalism introduced by Holland \cite{holland00,holland88}, where a quantum rigid $SU(2)$ rotor is represented by  an ensemble of $SO(3)$ rotors. 
The standard quantum mechanics does not describe coordinates and trajectories of particles, but only expectation values of various observables and the probabilities for particular outcomes of experiments. On the other hand, in the Bohmian interpretation of quantum mechanics, one can additionally discuss positions and momenta of particles and the probability densities that particles actually {\it are} at particular positions with a particular momentum, not only the probabilities that such positions or momenta {\it will be} obtained after completing the measurement process.

Following Bohm the wave function is expressed as 
\begin{equation}
\psi(\alpha,\beta,\gamma)=\mathcal{R} e^{i \mathcal{S}},
\label{psi}
\end{equation}
where $\mathcal{R}$ and $\mathcal{S}$ are real functions \cite{bohm1}. In this approach the angular momentum in the Bohmian space is given by a real, three dimensional vector ${\mathbf{M}}=i{\hat{\mathbf{M}}} \mathcal{S}$, 
\begin{eqnarray}
M_x&=&-\cos \beta \;\partial  \mathcal{S}/\partial \alpha +
\sin \beta \;\cot \alpha \;\partial  \mathcal{S}/\partial \beta
-{\sin \beta \over {\sin\alpha}} \;\partial  \mathcal{S}/\partial \gamma,\\
M_y&=&\sin \beta \;\partial  \mathcal{S}/\partial \alpha
+\cos \beta \;\cot \alpha \;\partial  \mathcal{S}/\partial \beta
-{\cos \beta \over \sin \alpha} \;\partial  \mathcal{S}/\partial \gamma,\\ 
M_z&=&-\partial  \mathcal{S}/\partial \beta.
\label{Mxyz}
\end{eqnarray}
The dynamics is determined from the Hamilton-Jacobi-type equations for the classical Hamiltonian with an additional quantum potential $Q$,
\begin{equation}
H={{\mathbf{M}}^2 \over 2 I}+Q,\quad Q={\hat{\mathbf{M}}^2 \mathcal{R}\over 2I \mathcal{R}}.
\label{hclass}
\end{equation}
The quantum potential $Q$ generates  the quantum torque $ \mathbf{T}=-i \hat{\mathbf{M}} Q$ acting on the rotor as
$d \mathbf{M}[\lambda(t)]/dt=\mathbf{T}[\lambda(t)]$. The  
equations of motion reduce to a set of first order non-linear differential equations for the trajectories in the configuration space, 
\begin{eqnarray}
I \dot{\alpha}&=&\partial  \mathcal{S}/\partial \alpha,  \label{a1}\\
I \dot{\beta}&=&(\partial  \mathcal{S}/\partial \beta
-\cos \alpha \;\partial  \mathcal{S}/\partial \gamma)/\sin^{\rm 2} \alpha, \label{b1}\\ 
I \dot{\gamma}&=&(\partial  \mathcal{S}/\partial \gamma
-\cos \alpha \;\partial  \mathcal{S}/\partial \beta)/\sin^{\rm 2} \alpha. \label{g1}
\end{eqnarray}
where particular trajectories $\lambda(t)$  in the space of hidden variables are the ensemble representatives, determined by the initial condition $\lambda(0)=\{\alpha_0,\beta_0,\gamma_0\}$.

\subsection{Spin-$\frac{1}{2}$}

In the analysis of  spin $s=\frac{1}{2}$  systems we chose solutions with spin projection along the rotor axis $n=\frac{1}{2}$ and along the $z$-axis $m=\pm\frac{1}{2}$, where the corresponding eigenstates are the Wigner matrices $D^s_{mn}$,
\begin{eqnarray}
u_\uparrow(\lambda)&=&{1 \over \sqrt{8\pi^2}}D_{{1 \over 2} {n}}^{s}(\beta,\alpha,\gamma)={1 \over \sqrt{8\pi^2}}e^{-i\beta/2-i n \gamma}\cos {\alpha \over 2},\\
u_\downarrow(\lambda)&=&{1 \over \sqrt{8\pi^2}}D_{-{1 \over 2} {n}}^{s}(\beta,\alpha,\gamma)={1 \over \sqrt{8\pi^2}}e^{i\beta/2 -i n\gamma}\sin {\alpha \over 2}.
\label{upm}
\end{eqnarray}
One should note that the core of this approach lies in the coexistence of the motion of a rigid top whose configuration space is $SO(3)$ but its motion is guided by a wave and the quantum potential whose spin configuration space is $SU(2)$, as discussed in Ref.~\cite{holland00}. Wave functions $\psi$ are normalized, and $\mathcal{R}^2$ represents the probability density for $\lambda$, with the normalization condition in the domain $\Lambda$
\begin{equation}
\int_0^{\pi}\int_0^{2\pi}\int_0^{4\pi} \mathcal{R}^2\sin\alpha\, d\alpha\, d \beta\, d \gamma\equiv \int_\Lambda \mathcal{R}^2 {\mathrm d}\lambda=1.
\label{norm}
\end{equation}
The quantum equilibrium ensemble average of some function $\mu(\lambda)$ is then given by
\begin{equation}
\langle \mu \rangle= \int_\Lambda \mu(\lambda)\mathcal{R}^2(\lambda) {\mathrm d}\lambda.
\label{ave}
\end{equation}

As an example, consider a qubit in the "spin up" state,  $\psi=\langle\lambda\left|\uparrow\right\rangle=u_\uparrow$. The exact solution  is given by
$\lambda(t)=\{\alpha_{0},\beta_{0}-t/\tau,\gamma_{0}-t/\tau\}$, where $\{\alpha_{0},\beta_{0},\gamma_{0}\}$ are the initial values for $\lambda$ and
$\tau=4 I \cos^2 (\alpha_{0}/2)$. The angular momentum  vector  precesses uniformly anticlockwise about the $z$-axis with a constant polar angle $\theta=\alpha_{0}/2$, 
\begin{eqnarray}
{\mathbf{M}}(t)&=&|{\mathbf{M}}|(\sin \theta \sin \phi,\sin \theta \cos \phi,\cos\theta),\\
|{\mathbf{M}}|&=&\frac{1}{2 \cos\theta},\quad\phi=\beta_{0}-t/\tau.
\label{mzeta}
\end{eqnarray}
Note that $M_z$ and ${\mathbf{e}}\cdot{\mathbf{M}}$
are constant $\frac{1}{2}$, while the angular momentum vector length is constant, $|{\mathbf{M}}|\geq\frac{1}{2} $, dependent of the initial condition for $\lambda(t)$.

\subsection{Entangled qubit pair}

The generalization to the case of two rotors is straight\-forward \cite{holland00}. We consider here a general two qubit   state with vanishing total angular momentum projection, Eq.~(\ref{psi2}). 
The guiding wave function $\psi(\lambda)=\langle\lambda|\Psi\rangle$,
\begin{equation}
\psi(\lambda)= \cos {\vartheta \over 2} u_\uparrow(\lambda_1)u_\downarrow(\lambda_2)+e^{i \varphi} \sin {\vartheta \over 2}u_\downarrow(\lambda_1) u_\uparrow(\lambda_2),
\label{psizeta}
\end{equation}
is given in six dimensional space spanned by $\lambda=\{\lambda_1,\lambda_2\}$, with $\lambda_{1,2}$ representing the coordinates of the first and the second rotor, respectively. The angular momenta  are determined by the  set of equations (\ref{psi}-\ref{Mxyz}) and (\ref{a1}-\ref{g1}), generalized to the case of two qubits, ${\mathbf{M}}_{1,2}=i{\hat{\mathbf{M}}_{1,2}} \mathcal{S}$.
The corresponding Hamiltonian is given by
\begin{equation}
H={{\mathbf{M}}_1^2+ {\mathbf{M}}_2^2  \over 2 I}+{(\hat{\mathbf{M}}_1^2 +\hat{\mathbf{M}}_2^2)\mathcal{R}\over 2I \mathcal{R}}.
\label{h12}
\end{equation}
It should be noted, that even for two non-interacting, but entangled qubits, the 
quantum potential $Q(\lambda)$ -- the second term in Eq.~(\ref{h12}) -- represents an instant {\it interaction} between the rotors as a fingerprint of the quantum nature of the problem. The solutions for each of the 
angular momentum vectors ${\mathbf{M}}_{1}$ and ${\mathbf{M}}_{2}$  are functions of six common coordinates forming
the trajectory $\lambda(t)$  determined by six initial values $\lambda(0)$.
The ensemble average is for the case of two qubits also given by Eq.~(\ref{ave}), but in six-dimensional space and with ${\mathrm d}\lambda={\mathrm d}\lambda_1 {\mathrm d}\lambda_2$. 

For the case of non-entangled qubits, $\vartheta=0$, the solutions are given by appropriately applied Eq.~(\ref{mzeta}), while for a general $\vartheta$ the angular momenta and the rotor axes are not coplanar, Fig.~\ref{fig1}(a), and exhibit a rich variety of precessional motions. Total energy  $E=s(s+1)/I$ is a constant of motion equal for all $\lambda$. 
Even though also some other constants of motion can readily be  identified and one can prove  that 
the orbits $\lambda(t)$ are periodic in particular subspaces of the phase space, the time evolution along the orbits can not be tracked 
analytically in general.  The analysis of the topology of the orbits reveals several classes forming distinct portions of the phase space and the survey of these properties of the model will be presented elsewhere \cite{dBBdinamika}.

\section{The probability distributions}

Here we focus on the analysis of various static, $t=0$,  probability distributions of two spin-$\frac{1}{2}$ rotors and we compare
particular ensemble averages  with the corresponding results obtained by standard quantum mechanics. 
All probability distributions and the ensemble averages considered here are time-independent while some quantities specific for time-dependent properties of the rotors  are presented in Ref.~\cite{epl11}. 

\subsection{Single particle properties}

First we consider single particle properties of the rotor pair. For the state $|\Psi\rangle$ with a vanishing total spin projection the equality  $M_{1z}(\lambda)+M_{2z}(\lambda)=0$ is fulfilled for all $\lambda$ and the projections along the rotor axes are constants of motion by construction, ${\mathbf{e}}_{1,2}\cdot{\mathbf{M}}_{1,2}=\frac{1}{2}$.
The average values of the angular momentum calculated by Eq.~(\ref{ave}) coincide with the standard result for spin expectation values 
\begin{equation}
\langle {\mathbf{M}}_{1} \rangle=-\langle {\mathbf{M}}_{2} \rangle =\frac{1}{2}(0,0,\cos \vartheta)=\langle \Psi|\mathbf S_1|\Psi\rangle,
\label{ms}
\end{equation}
where  $\mathbf{S}_1={1\over2}(\sigma_{1x},\sigma_{1y},\sigma_{1z})$ is the ordinary spin operator for the first particle and $\sigma$ are the Pauli matrices. 

Despite that average Bohmian momentum $\mathbf{M}_1$ and standard quantum counterpart $\mathbf{S}_1$ are equal in this case, there is a significant difference of both angular momentum vectors: the state $\left|\Psi\right\rangle$ is an eigenstate of $\mathbf{S}_1^2$, which is dispersionless, while $\left|\mathbf{M}_1\right |^2$ is different for different members of the ensemble and distributed according to some probability distribution corresponding to the relation Eq.~(\ref{mzeta}).
\begin{figure}[htbp]
\begin{center}
\includegraphics[width=70 mm]{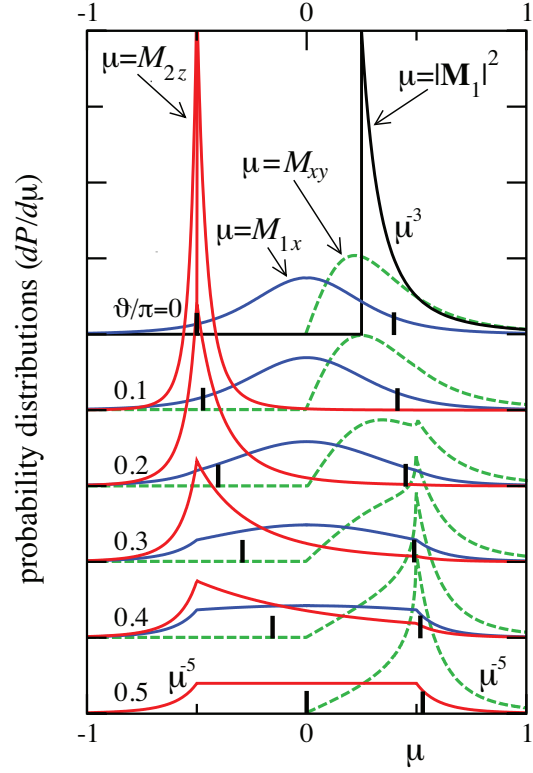}
\caption{The probability distribution $dP/d\mu$ of $\mu= M_{1x}$ for various $\vartheta$ (blue lines). For unentangled qubits  it is given by $dP/d\mu=\frac{3}{2}(1+4\mu^2)^{-5/2}$ while the distribution for $\mu= M_{2z}$ (red lines) is  for $\vartheta=0$ a delta-function at $\mu=-\frac{1}{2}$.
The distributions are independent of $\varphi$ and are  shifted vertically for clarity. For
$\mu=|\mathbf{M}_{1}|^2$  the distribution is given analytically by $dP/d\mu=\Theta(\mu-\frac{1}{4})(2\mu)^{-3}$ [derived from Eq.~(\ref{dpdm})] and is independent of $\vartheta$ (black line).
The probability distribution of the $xy$-plane projection of $\mathbf{M}_1$, $\mu=\sqrt{M_{1x}^2+M_{1y}^2}= M_{xy}$, is shown with green dashed lines. Thick vertical lines represent  $\langle M_{2z} \rangle$ (left set) and $\langle M_{xy} \rangle$ (right set).}
\label{fig2}
\end{center}
\end{figure}
In order to gain the insight into the properties of the Bohmian motion of the rotors we
analyse this and various other probability distributions corresponding to hidden variables. In general the quantum equilibrium probability distributions corresponding to various functions $\mu(\lambda)$ studied here are defined by
\begin{equation} 
{dP(\mu) \over d\mu}=\int_\Lambda \delta[\mu-\mu(\lambda)]\mathcal{R}^2(\lambda) {\mathrm d}\lambda,
\label{dpdmu}
\end{equation}
where $\delta(\mu)$ is the Dirac delta function. 
Average values corresponding to such distributions are then expressed, {\it e.g.}, for $\mu(\lambda)= M_{1x}$, by
\begin{equation}
\langle M_{1x} \rangle=\int_\Lambda M_{1x}(\lambda)\mathcal{R}^2(\lambda)\mathrm{d} \lambda=\int_{-\infty}^\infty M_{1x}{d P\over d M_{1x}}d M_{1x}.
\end{equation}
Therefore $d P$ represents the probability that for a given qubit pair in the state $|\Psi\rangle$ the $x$-axis angular momentum  component of the first particle {\it is} within the interval $[M_{1x},M_{1x}+dM_{1x}]$ not the probability that $M_{1x}$ {\it will} be obtained after the performed measurement. Note that the two probability densities are in general not equal, while the probability for the outcome of a particular experiment is always identical for both approaches, the Bohm and the standard quantum mechanics.

For few limiting cases the probability distributions can be derived analytically, while in general one has to rely on a numerical evaluation of  Eq.~(\ref{dpdmu}), for example, by the limit of a sum over the configuration space, 
\begin{equation}
{dP(\mu) \over d\mu}=\lim_{N\to \infty \atop \epsilon\to 0} {\sum_{i=1}^{i=N} \delta_\epsilon[\mu-\mu(\lambda_i)]\mathcal{R}^2(\lambda_i)
/ \sum_{i=1}^{i=N} \mathcal{R}^2(\lambda_i)},
\label{num}
\end{equation}
where $\delta_\epsilon(\mu)$ is a normalized rectangular function of width $\epsilon$ and $N$ is the total number of samples 
$\lambda_i$ forming a uniform grid  in the configuration space spanned by $\{\cos\alpha_1,\beta_1,\gamma_1,\cos\alpha_2,\beta_2,\gamma_2\}$.
The hidden variables corresponding to the internal rotation around the  principal axes of the rotors are irrelevant for the quantities considered here, thus we fixed $\gamma_{1,2}\equiv0$. To obtain sufficiently converged results  with Eq.~(\ref{num}) we used  $N=1024^4$ and $\epsilon=10^{-3}$ for the full domain $\Lambda$ corresponding to $|\cos \alpha_{1,2}|\leq1$ and $0\leq\beta_{1,2}< 2\pi$ while for particular cases with higher symmetry the domain can be reduced.
The error due to the finite grid used in the calculations  presented throughout the paper is always below the width of the lines in the corresponding figures.

The probability distribution of momentum length ${d P/ d |\mathbf{M}_{1}|}$ and the corresponding moments
can be for $\vartheta=0$ derived analytically, 
\begin{eqnarray}
{d P\over d |\mathbf{M}_{1}|}&=&{\Theta(|\mathbf{M}_{1}|-\frac{1}{2})\over 4|\mathbf{M}_{1}|^{5}}, \label{dpdm}\\
\left\langle |\mathbf{M}_{1}|^2\right\rangle&=&\frac{1}{2}=\frac{2}{3}\left\langle \Psi|\mathbf{S}_{1}^2|\Psi\right\rangle, \label{dpdm2}
\end{eqnarray}
where $\Theta(\mu)$ is the Heaviside step function. 

All single particle distributions are independent of $\varphi$,  and
the numerical analysis gives a firm evidence that the probability distribution Eq.~(\ref{dpdm}) is  in fact also independent of $\vartheta$. Numerically we determined also the probability distribution for the $xy$-plane projection of
the angular momentum, $\mu=\sqrt{M_{1x}^2+M_{1y}^2}\equiv M_{xy}$, for unentangled qubits given analytically by 
$dP/d\mu=16 \mu(1+4\mu^2)^{-3}$.
The distribution is shown in Fig.~\ref{fig2} for various $\vartheta$ and 
the average $\langle M_{xy} \rangle$,  right set of thick vertical lines in Fig.~\ref{fig2}, ranges from $\frac{\pi}{8}$ to $\frac{\pi}{6}$ from the unentangled to the fully entangled case, respectively, which is less than the standard quantum mechanics counterpart $\langle \Psi|\sqrt{S_{1x}^2+S_{1y}^2}|\Psi\rangle = \frac{1}{\sqrt{2}}$. 

Additionally are shown the probability distributions for $\mu= M_{1x}$ and $\mu= M_{2z}=-M_{1z}$. The average 
$\langle M_{2z} \rangle=-\frac{1}{2}\cos\vartheta$ is shown by the left set of thick vertical lines. The corresponding probability distribution progressively develops from a delta-function for unentangled qubits into a distinctive probability distribution curve, constant for  $|\mu|\leq\frac{1}{2}$,
for full entanglement at $\vartheta=\pi/2$. The characteristic kinks in the distributions at $|\mu|=\frac{1}{2}$ have the origin in Eq.~(\ref{edotm}) with the eigenvalue $n=\frac{1}{2}$ which sets the lower limit for the magnitude of the angular momentum in Eq.~(\ref{dpdm}).  

For perfectly entangled qubit pairs
the probability distribution of  the polar angle  $\cos \theta_1=M_{1z}/|\mathbf{M}_1|$ [Fig.~\ref{fig1}(a)] is constant,
$dP/d \cos \theta_1=\frac{1}{2}$, as anticipated for a singlet (but valid  for any $\varphi$, as argued above).  As a direct consequence emerge the exact expressions for the probability distributions of  $\mu={M}_{1z}$, $\mu={M}_{1z}^2$ (identical are the results corresponding to $\mu={M}_{1x(y)}$) and $\mu=M_{xy}$,
\begin{eqnarray}
{d P\over d \mu}&=&\frac{4}{5}\mathrm{min}(1,|2\mu|^{-5} ), \quad \mathrm{for} \;\mu={M}_{1z},  \label{dpdmz}\\ 
{d P\over d \mu}&=&\frac{8}{5}\mathrm{min}[(4\mu)^{-1/2} ,(4\mu)^{-3} ], \quad \mathrm{for} \;\mu={M}_{1z}^2, \label{dpdmz2}\\
{d P\over d \mu}&=&{2\over  15 \mu^5}[1-\Theta(\frac{1}{2}-\mu)(1+2\mu^2+6\mu^4)\sqrt{1-4\mu^2}], \quad \mathrm{for} \;\mu=M_{xy}.\label{dpdMxy2}
\end{eqnarray}

Finally, let us comment on an interesting property following from Eq.~(\ref{dpdm2}). The average square of the Bohmian momentum represents exactly $\frac{2}{3}$ of the corresponding quantum value, which can be by taking into account 
the equality of the Bohmian average energy and the quantum expectation value, $E=\langle H \rangle=\langle\Psi| \hat H |\Psi\rangle$, re-expressed into the identity
\begin{equation}
\left\langle{{\hat{\mathbf{M}}_{1,2}^2 \mathcal{R}\over \mathcal{R}}}\right\rangle=\frac{1}{2}\left\langle |\mathbf{M}_{1,2}|^2 \right\rangle.
\label{virial}
\end{equation}
The identity represents the virial theorem: 
for an arbitrarily entangled, but non-interacting, two qubit state the ensemble average of the quantum potential in Eq.~(\ref{h12}) represents exactly one half of the average kinetic energy in the Bohm space of hidden variables, 
\begin{equation}
\left\langle Q\right\rangle=\frac{1}{2}\left\langle H-Q\right\rangle.\label{virial}
\end{equation}
This means that the quantum potential contribution is never negligible, $\left\langle Q\right\rangle=E/3$, which indicates the contextual nature of considered hidden-variables formulation of quantum mechanics \cite{bell87}, {\it i.e.},  various angular momentum correlation functions after the performed measurement are in general different from Bohmian pre-measurement values where the rotational energy is diminished on the account of the potential part. This is
similar to the more familiar case of the orbital motion of point-like particles where the Bohmian particle velocity correlations differ from the canonical momentum quantum counterparts \cite{bohm2}. 

\subsection{Entanglement of formation and the Shannon entropy}

Apart from the ontological point of view,
an important distinction between standard and de Broglie-Bohm interpretation of quantum mechanics reflects in the probability distributions which can be discrete in one  case and continuous in the other \cite{komentar1}. Due to this qualitatively different nature of the probabilities the entanglement of a qubit pair in the two approaches is quantified differently.

The probability distribution for  a measurement outcome of spin projection  $S_{1z}$ (or $\hat{M}_{1z}$)  is discrete with the probabilities
$p_1=\cos^2{\frac{\vartheta}{2}}$ and $p_2=1-p_1$ for  "spin up" and "spin down", respectively. An appropriate measure for the degree of entanglement  is 
the entanglement of formation \cite{bennett}, the asymptotic conversion rate to maximally entangled states from an ensemble of  copies of a non-maximally entangled state \cite{vedral}. It is given by
the von Neumann entropy $S(\rho_1)=-{\rm Tr} \rho_1 \log \rho_1$, where $\rho_1$ is the reduced density matrix obtained by tracing the qubit pair density matrix  over the  degrees of freedom of particle 2. The log is to the base 2 
and entropy is expressed in bits. In our case of a pure state with $\rho=\left| \Psi\right\rangle\left\langle \Psi\right |$ the von Neumann entropy is identical to the binary (Shannon) entropy 
\begin{equation}
H_S=-\sum_i p_i \log p_i,
\label{shannon}
\end{equation}
where is $i=1,2$ for $S_{1z}=\pm\frac{1}{2}$, respectively.

\begin{figure}[tbhp]
\begin{center}
\includegraphics[width=75mm]{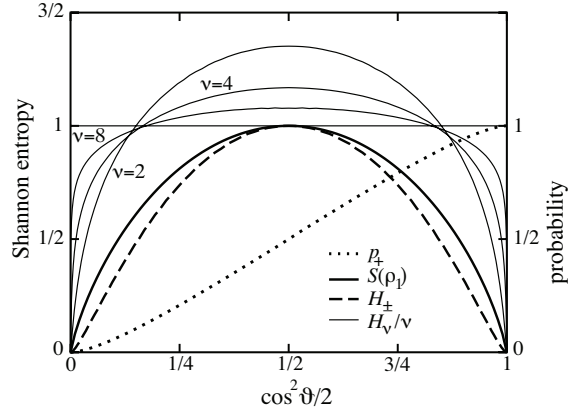}
\caption{Probability $p_+$ for $M_{1z}\geq0$ as a function of $p_1=\cos^2{\frac{\vartheta}{2}}$ (dotted line, right scale). Note the deviation from a straight line which reflects in the inequality of the binary entropy $H_\pm$ (dashed) and the entanglement of formation $S(\rho_1)$ (thick full line). Thin lines represent the renormalized Shannon entropy $H_\nu/\nu$  corresponding to bin widths $\Delta=2^{-\nu}=\frac{1}{4}, \frac{1}{16}, \frac{1}{256}$.}
\label{fig3}
\end{center}
\end{figure}

In de Broglie-Bohm approach $\mu=M_{1z}$ is a continuous random variable thus the probability distribution $p(\mu)=d P/ d \mu$ is not discrete, as discussed in the previous subsection.  In this case
an analogue measure of entanglement is the Shannon entropy $H_\pm$ for a random binary variable that 
the angular momentum vector of the first particle is in the upper or lower hemisphere with
probabilities  $p_+=\int_0^\infty p(\mu) d \mu$ and $p_-=1-p_+$, respectively. Although  $\left\langle M_{1z} \right \rangle=\left\langle S_{1z} \right \rangle$, Eq.~(\ref{ms}), the probabilities $p_1$ and $p_+$ are in general not equal as is shown  in Fig.~\ref{fig3} (dotted line), where are  as a function of $p_1$ shown also the binary entropy $H_\pm$ (dashed) and the entanglement of formation (thick full line).

Since the probability distribution $p(M_{1z})$ is a continuous function an appropriate measure is given also by the differential entropy,
\begin{equation}
h=-\int_{-\infty}^\infty p(\mu) \log p(\mu)d \mu.
\label{diffen}
\end{equation}
The differential entropy  is related to the Shannon entropy Eq.~(\ref{shannon}) with quantized probability  $p_i=p(\mu_i) \Delta$, where the range of $\mu$ is divided into bins of length $\Delta$ and $\mu_i$ is a value of $\mu$ within the $i$-th bin such that $p_i=\int_{(i-\frac{1}{2})\Delta}^{(i+\frac{1}{2})\Delta} p({\mu})d \mu$  \cite{cover}. For the sake of convenience we take $\Delta=2^{-\nu}$ which leads to a relation valid for  the corresponding Shannon entropy $H_\nu$, 
\begin{equation}
H_\nu -\nu \to h,\quad{\rm as}\;\nu\to\infty.
\label{hnu}
\end{equation}
This means that $\nu+h$ is approximately the number  of bits on the average required to describe $M_{1z}$ to $\nu$ bit accuracy. The entropy $H_\nu$ is zero for unentangled qubits and is maximum for a fully entangled case. 
Maximum of $h$, calculated from Eq.~(\ref{dpdmz}), equals $h_{max}=\log\frac{5}{4}e^{1/4}\approx0.68$  with the corresponding support length containing most of the probability ${\rm Vol}( M_{1z})=2^{h_{max}}\approx1.61$, which can also be estimated visually from Fig.~\ref{fig2}.
In Fig.~\ref{fig3} is with thin lines presented normalized Shannon entropy $H_\nu/\nu$ for $\nu=2,4,8$.  We found an excellent agreement with the asymptotic relation  Eq.~(\ref{hnu}) for $\nu\gtrsim6$ where  the deviations are below the width of the lines.

The probability distribution $d P/d M_{1z}$ thus determines the degree of entanglement similarly to the conventional approach if binary entropy criterion is used. Quantitative differences between $H_\pm$ and $S(\rho_1)$ vanish for unentangled or fully entangled pairs, as expected.
The main difference between de Broglie-Bohm  and the standard view is, however, that the information contained in de Broglie-Bohm probability distribution is due to its continuous nature dependent on the discretization, therefore can be arbitrarily large, $\sim \nu+h$.  It should be noted that all this information is not accessible and probably can not be used for a transfer of information because during the process of spin measurement the spin projection outcomes will become $\pm\frac{1}{2}$ -- {\it i.e.}, a non-continuous value -- thus the entanglement of formation is limited to at most one bit per entangled qubit pair.

\subsection{Two-particle distributions}

\begin{figure}[htbp]
\begin{center}
\includegraphics[width=70 mm]{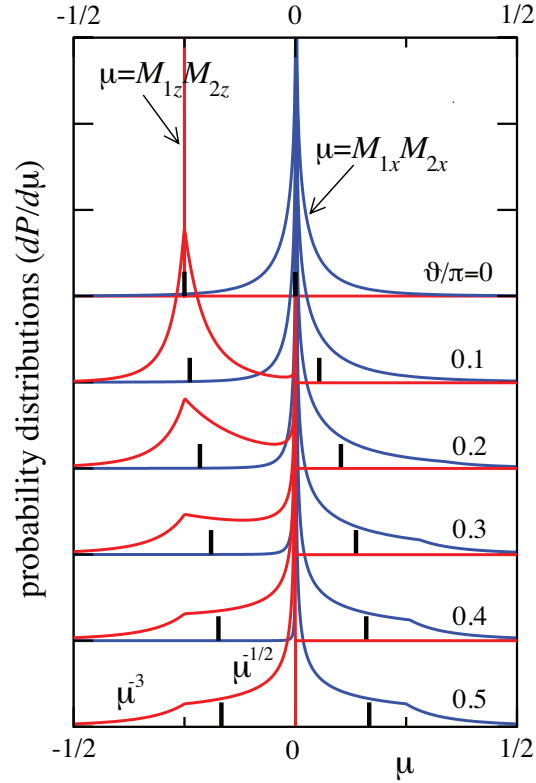}
\caption{Probability distributions $dP/d\mu$ for $\mu= M_{1x}M_{2x}$ (blue lines) and 
$\mu= M_{1z}M_{2z}$ (red lines) for various $\vartheta$ and $\varphi=0$. The distributions are vertically shifted for clarity.  Thick vertical lines represent  $\langle M_{1z}M_{2z} \rangle$ and $\langle M_{1x}M_{2x} \rangle$ (left and right set, respectively).
}
\label{fig4}
\end{center}
\end{figure}

One of the main motivations of the present paper is the analysis of the two-particle properties which reflect quantum mechanical correlations of two entangled qubits. We consider here the probability distributions $dP/d\mu$ for $\mu= M_{1x}M_{2x}$ and $\mu= M_{1z}M_{2z}$. The probability distributions are presented in Fig.~\ref{fig4} for various degrees of entanglement and for $\varphi=0$.  

The distribution of $M_{1z}M_{2z}$ is independent of $\varphi$ and is for
two unentangled qubits a delta function  at $\mu=-\frac{1}{4}$ while for fully entangled triplet state  it takes a form identical to the distribution of $M_{1x}M_{2x}$ but with opposite  sign of $\mu$. 
For the singlet the two distributions are identical (not shown). 
Due to the equality $M_{1z}=-M_{2z}$  one can for $\vartheta=\pi/2$  from  Eq.~(\ref{dpdmz2})  express these distributions  analytically,
\begin{eqnarray}
{d P\over d \mu}&=&\frac{8}{5}\mathrm{min}[(4\eta\mu)^{-1/2} ,(4\eta\mu)^{-3} ], \quad \mathrm{for}\,\,
\mu=M_{1z}M_{2z},\label{dpdm12z}\\
{d P\over d \mu}&=&(4\eta\mu)^{-1/2}, \quad\mathrm{for}\,\,
\mu=M_{1z}M_{2z}/(|\mathbf{M}_1||\mathbf{M}_2|),\label{dpB}
\end{eqnarray}
where $\eta=-1$,  while  the  corresponding distributions of the $x$- and $y$-components of the angular momenta have the same form for  the singlet or the triplet, with  $\eta=\mp 1$, respectively. Here the kinks in the distributions are at $|\mu|=\frac{1}{4}$. 
Note that $\mu^{-1/2}$ in Eq.~(\ref{dpB}) reflects the isotropic distribution of the angular momentum components. Let us also point out that for every $\lambda$ the angular momentum  vectors have equal  length, $|\mathbf{M}_{1}(\lambda)|=|\mathbf{M}_{2}(\lambda)|$, while for  $\vartheta<\frac{\pi}{2}$
this equality holds  only on average, $\langle |\mathbf{M}_{1}|\rangle=\langle|\mathbf{M}_{2}|\rangle$, {\it i.e.}, the width of the probability distribution for $\mu=|\mathbf{M}_{1}|-|\mathbf{M}_{2}|$ is finite, for example, $\langle \mu^2\rangle=\frac{1}{9}$ is maximum at  $\vartheta=0$.

\begin{figure}[htbp]
\begin{center}
\includegraphics[width=70mm]{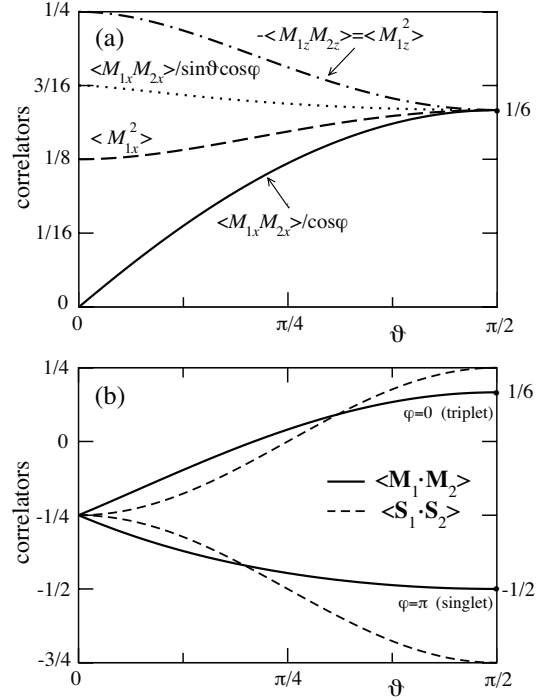}
\caption{(a) Renormalized spin-spin correlators:
$\langle M_{1x} M_{2x} \rangle/\cos \varphi$ or $\langle M_{1y} M_{2x} \rangle/\sin \varphi$ (thick full line). $\langle M_{1x} M_{2x} \rangle/\cos\vartheta\cos\varphi$ (dotted line), and  $\varphi$-independent $\langle M_{1z} M_{2z} \rangle=-\langle M_{1z}^2 \rangle$ (dashed-dotted line) and $\langle M_{1x}^2 \rangle=\langle M_{1y}^2 \rangle$ (dashed line). 
(b) Angular momentum correlator  $\langle \mathbf{M}_{1}  \cdot\mathbf{M}_{2} \rangle$ (full lines) in comparison with standard quantum mechanics  result 
$\langle \mathbf{S}_{1}\cdot  \mathbf{S}_{2} \rangle$  (dashed lines) for two values $\varphi=0$ and $\varphi=\pi$.}
\label{fig5}
\end{center}
\end{figure}

The ensemble average values $\langle M_{1z}M_{2z} \rangle$ and $\langle M_{1x}M_{2x} \rangle$ are indicated in Fig.~\ref{fig4} by thick vertical lines.
In general, the $\varphi$-dependence of correlators $\langle M_{1i}M_{2j}\rangle$ is identical to the standard quantum mechanical spin-spin correlation tensor,
\begin{eqnarray}
\langle\Psi| S_{1i} S_{2j}|\Psi \rangle={ 1 \over 4}
\left(
\begin{array}{lll} \sin \vartheta \cos \varphi & 
- \sin \vartheta \sin \varphi  &0\\
 \sin \vartheta \sin \varphi  &\sin \vartheta \cos \varphi &0 \\
0&0&-1 \\
\end{array}
\right),  
\label{sisj}
\end{eqnarray}
where $i(j)$ represent the corresponding $x,y,z$ components. In Fig.~\ref{fig5}(a) are shown various spin-spin correlators 
$\langle M_{1i}M_{2j} \rangle$ as a function of $\vartheta$.
The dependence of $\vartheta$ is slightly different in comparison with Eq.~(\ref{sisj}) but at $\vartheta=\frac{\pi}{2}$ the Bohmian correlators are exactly $\frac{2}{3}$ of their quantum counterparts, as is the case for the single-particle averages discussed in the previous subsection, Eq.~(\ref{dpdm2}). The second moments $\langle M_{1z}^2 \rangle$ and $\langle M_{1x}^2 \rangle$ are also shown for a comparison with standard quantum mechanics result $\langle S_{1x,y,z}^2 \rangle=\frac{1}{4}$.

In Fig.~\ref{fig5}(b) is with full lines presented the correlator $\langle {\mathbf{M}}_1\cdot   {\mathbf{M}}_2 \rangle$ for singlet and triplet-type of qubit pair configuration in comparison with quantum mechanical value given by $\langle\Psi| \mathbf{S}_{1}\cdot  \mathbf{S}_{2}|\Psi \rangle=
{1 \over 4}(2\sin\vartheta\cos\varphi-1)$ \cite{ramsakQM}. For unentangled state all correlators are equal to $-{1\over 4}$ while for $\vartheta=\pi/2$ the Bohmian correlators for any $\varphi$ exhibit anticipated relation  $\langle {\mathbf{M}}_1\cdot   {\mathbf{M}}_2 \rangle=\frac{2}{3}\langle\Psi| \mathbf{S}_{1}\cdot  \mathbf{S}_{2}|\Psi \rangle$ which together with Eq.~(\ref{dpdm2}) leads to the identity
\begin{equation}
\left\langle ({\mathbf{M}}_1+{\mathbf{M}}_2)^2 \right\rangle=\frac{2}{3}\langle\Psi| (\mathbf{S}_{1}+  \mathbf{S}_{2})^2|\Psi \rangle.
\label{sum}
\end{equation}
 
\section{Geometrical view of a qubit pair}

In standard quantum mechanics a visualization of a relative motion of two angular momenta of an entangled qubit pair is not given directly, although by appropriately defined angle operators some imagery can be given \cite{ramsakQM}. On the other hand, in de Broglie-Bohm interpretation of quantum mechanics an analysis  of  geometrical quantities as are, for example, relative angles between the momenta in the space of hidden variables is simple and directly feasible.
We apply the approach introduced in the previous Section and systematically present geometric quantities relevant to the qubit pair given by Eq.~(\ref{psi2}). We consider two possibilities:  the angle made by  the angular momenta and the azimuthal angle made by the $xy$-plane projections of the momenta, as follows.

\subsection{Angle between two angular momenta}

The angular momenta $\mathbf{M}_{1}$ and $\mathbf{M}_{2}$  of two qubits make  an angle $\Phi$ in the Bohmian space of hidden variables $\lambda$   as shown in Fig.~\ref{fig1}(b). The average $\cos \Phi$ and the corresponding dispersion $\Delta \cos \Phi$ are  given by the expressions
\begin{eqnarray}
\langle \cos \Phi\rangle&=&\left\langle {{ \mathbf{M}_1}\cdot{ \mathbf{M}_2} \over |{ \mathbf{M}_1}||{ \mathbf{M}_2}|}\right\rangle, \label{cosp}\\ 
(\Delta \cos \Phi)^2&=&\langle \cos^2 \Phi\rangle - \langle \cos \Phi\rangle^2.
\label{cospp}
\end{eqnarray}

\begin{figure}[htbp]
\begin{center}
\includegraphics[width=70mm]{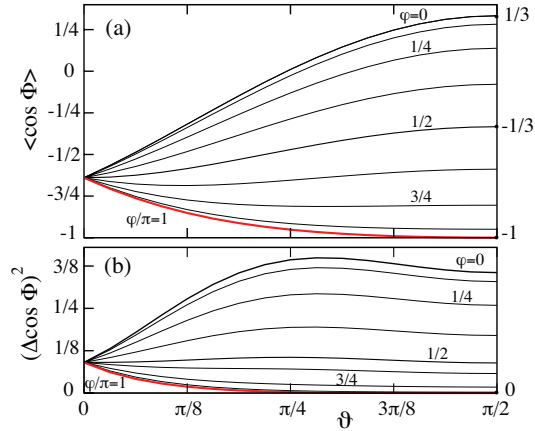}
\caption{(a) $\langle\cos \Phi\rangle$ vs. $\vartheta$  for various $\varphi$. Bullets represent exactly determined special cases. (b) Variances $\Delta \cos \Phi$ corresponding to (a). 
}
\label{fig6}
\end{center}
\end{figure}

The average cosine is presented in Fig.~\ref{fig6}(a) as a function of $\vartheta$ and for various $\varphi$. 
For $\vartheta=\pi/2$ we find exact relation $\langle\cos \Phi\rangle=\frac{1}{3}(2\cos\varphi -1)$,  identical to the standard quantum mechanics expression \cite{ramsakQM}.
As expected, perfect anti-parallel alignment is found for the singlet state ($\varphi=\pi$, red line), while momenta for the triplet state, $\varphi=0$, are only partially aligned,  $\langle\cos \Phi\rangle=\frac{1}{3}$. In Fig.~\ref{fig6}(b) is shown the variance $\Delta \cos \Phi$.  For $\varphi \sim \pi$ it is strongly suppressed for a wide range of $\vartheta\gtrsim\frac{\pi}{4}$ indicating a significant stability of anti-parallel angular momenta configuration of the singlet qubit pairs (red line). The variance for states close to the triplet configuration, on the other hand, is much higher and is of the order of average  $\cos \Phi$. 

\begin{figure}[htbp]
\begin{center}
\includegraphics[width=70mm]{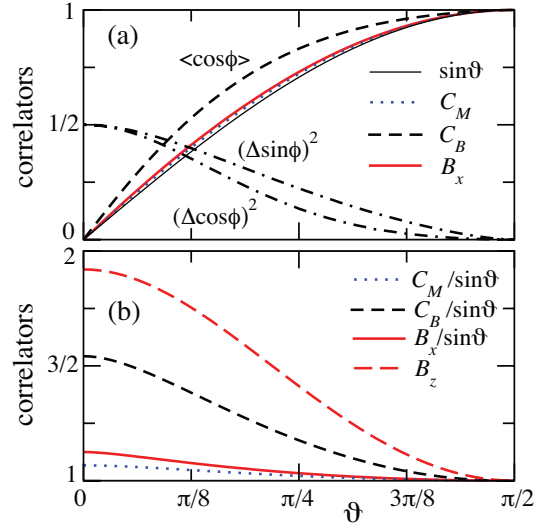}
\caption{$\sin\vartheta$ (thin line), $C_M$ (dotted line), $C_B$ (equal to $\langle \cos \phi\rangle$ for  $\varphi=0$, dashed line), and $B_{x}(\vartheta)$ (red thick full line).  Variances  $(\Delta \cos \phi)^2$ and $(\Delta \sin \phi)^2$ are shown for $\varphi=0$ or $\pi$ 
(dashed-dotted lines).  For $\varphi=\pm\pi/2$ the two curves are interchanged. $B_{z}(\vartheta)$ (red long dashed line) and the same set of quantities as in (a), but renormalized by $\sin \vartheta$. Note the identity of all curves at $\vartheta=\frac{\pi}{2}$.
}
\label{fig7}
\end{center}
\end{figure}

The correlators applied in Eq.~(\ref{cosp},\ref{cospp}) exhibit a similar symmetry structure as $\langle\Psi| S_{1i} S_{2j}|\Psi \rangle$ in Eq.~(\ref{sisj}),
\begin{eqnarray}
\left\langle  {M_{1i}M_{2j} \over |\mathbf{M}_1| |\mathbf{M}_2|}\right\rangle={ 1 \over 3}
\left(
\begin{array}{lll} B_x \cos \varphi & 
- B_x \sin \varphi  &0\\
 B_x \sin \varphi  &B_x \cos \varphi &0 \\
0&0&-B_z \\
\end{array}
\right),  
\label{bij}
\end{eqnarray}
where $B_{x,z}$ are functions of $\vartheta$, shown in  Fig.~\ref{fig7}(a,b) (red lines) and $\langle \cos \Phi\rangle=\frac{1}{3}[2B_x(\vartheta) \cos\varphi-B_z(\vartheta)]$. 

For comparison is in Fig.~\ref{fig7}(a,b)
additionally shown the Bohmian counterpart of $\sin \vartheta$ in Eq.~(\ref{sisj}), $C_M=6\langle M_{1x} M_{2x} \rangle/\cos\varphi=6\langle M_{1y} M_{2x} \rangle/\sin\varphi$ (dotted line). Let us remark that both, $B_x$ and $C_M$, are very close,  though not identical to $\sin \vartheta$.

\subsection{Relative azimuthal angle}

We consider here qubit pairs with vanishing total $z$-axis projection of spin, therefore the angles between projections of $\mathbf{M}_{1,2}$ onto the $xy$-plane play a special role \cite{epl11}. We analyze the difference of azimuthal angles of the angular momenta $\phi=\phi_2-\phi_1$ via properties of average cosine and the corresponding variance defined by 
\begin{eqnarray}
\left\langle \cos \phi\right\rangle&=&\left\langle {{M_{1x} M_{2x}+M_{1y} M_{2y}} \over  
\sqrt{(M_{1x}^2 +M_{1y}^2)(M_{2x}^2 +M_{2y}^2)} }\right\rangle, \\ 
(\Delta \cos \phi)^2&=&\langle \cos^2 \phi\rangle - \langle \cos \phi\rangle^2.
\label{fixy}
\end{eqnarray}
By inspection of the wave function $\psi$, Eq.~(\ref{psizeta}),  we find that a finite $\varphi$ represents only a shift 
of one of the azimuthal angles for each member of the ensemble, $\beta_2\to \beta_2+\varphi$, which results in the identity $\langle \phi \rangle=\varphi$ and the
decoupling $\langle \cos \phi\rangle=C_B\cos\varphi$, where $C_B$ is a function of $\vartheta$ only.  An analogous result holds for $\langle \sin \phi\rangle=C_B\sin\varphi$, thus $\langle \cos (\phi-\varphi)\rangle=C_B$ and $\langle \sin (\phi-\varphi)\rangle=0$. In Fig.~\ref{fig7}(a)  is $C_B(\vartheta)$  presented by a dashed line in comparison with $\sin\vartheta$ (thin full line).
The azimuthal angle fluctuations can be analyzed by variances $\Delta \cos \phi$ and $\Delta \sin \phi$,  also shown in Fig.~\ref{fig7}(a) (dashed-dotted lines).  Note that with increasing $\vartheta$ 
suppressed angle fluctuations signal 
a higher degree of entanglement as a highly correlated distribution of angular momenta making azimuthal angles progressively closer to $\varphi$.

\section{Bell's inequalities for a Bohmian ensemble}

One of the most striking properties of two entangled qubits is the violation of Bell's inequality \cite{bell64} derived for systems described by local deterministic hidden-variable theories. For more general versions of the inequality, Clauser, Horne, Shimony, and Holt \cite{chsh69} showed that
\begin{equation}
|P(\mathbf{a},\mathbf{b})+P(\mathbf{a},\mathbf{b}')+P(\mathbf{a}',\mathbf{b})-P(\mathbf{a}',\mathbf{b}')|\leq 2,
\label{chsh}
\end{equation}
where $P$ is the counting correlation of measured outcomes of spin projections along polarizer directions given by normalized vectors $\mathbf{a}$, $\mathbf{b}$,
$\mathbf{a}'$ and $\mathbf{b}'$ \cite{bell64}. It is readily seen that for the quantum prediction
\begin{equation}
P(\mathbf{a},\mathbf{b})=\langle\mathrm{singlet}| (\mathbf{a} \cdot \vec\sigma_1)( \mathbf{b}\cdot \vec\sigma_2 )|\mathrm{singlet}\rangle=-\mathbf{a}\cdot\mathbf{b}
\label{c0}
\end{equation}
with well chosen polarizer directions the inequality can be violated by a factor as large as $\sqrt{2}$.

Theory of two entangled qubits presented here is based on  the hidden-variable space spanned by $\lambda$, but the formalism is non-local due to the quantum potential generating instantaneous interaction between the rotors. The question arises then to which degree the inequality Eq.~(\ref{chsh}) is respected for the case of two rotors represented as a Bohmian ensemble.

Let us first reconsider spin-spin correlations 
$C(\mathbf{a},\mathbf{b})=\langle\Psi| (\mathbf{a} \cdot \vec\sigma_1)( \mathbf{b}\cdot \vec\sigma_2 )|\Psi\rangle$
for a general state $|\Psi\rangle$, Eq.~(\ref{psi2}). Correlations can be for convenience expressed directly in terms of spin operators $\mathbf{S}_{1,2}$ and $\mathbf{S}_{1,2}^2$ as,
\begin{eqnarray}
C(\mathbf{a},\mathbf{b})
&=&3\langle\Psi|{ (\mathbf{a} \cdot \mathbf{S}_1)( \mathbf{b}\cdot \mathbf{S}_2 )\over \sqrt{\mathbf{S}_1^2 \mathbf{S}_2^2}}|\Psi\rangle
=\label{cb}\\
&=&(a_x b_x+a_y b_y)\sin\vartheta\cos\varphi+ (a_y b_x-a_x b_y)\sin\vartheta\sin\varphi-a_z b_z.\nonumber\\
\nonumber
\end{eqnarray}
In a similar manner
one can construct for the case of Bohmian rotors an analogous correlator, related to the quantities $B_{x,z}$ studied in the previous section,
\begin{eqnarray}
B(\mathbf{a},\mathbf{b})&=&3\left\langle{ (\mathbf{a} \cdot \mathbf{M}_1)( \mathbf{b}\cdot \mathbf{M}_2 )\over |\mathbf{M}_1| |\mathbf{M}_2|}\right\rangle=\label{bb}\\
&=&(a_x b_x+a_y b_y)B_{x}\cos\varphi+(a_y b_x-a_x b_y)B_{x}\sin\varphi-a_z b_z B_{z}.\nonumber\\
\nonumber
\end{eqnarray}
For a perfectly entangled qubit pair
\begin{equation}
\left|\Psi\right\rangle=\frac{1}{\sqrt{2}}(\left|\uparrow\downarrow\right\rangle+e^{i \varphi} \left|\downarrow\uparrow\right\rangle),
\label{epr}
\end{equation}
the correlations $C(\mathbf{a},\mathbf{b})$ and $B(\mathbf{a},\mathbf{b})$ are identical due to the equality $B_{x}=B_z=1$, see Fig.~\ref{fig7}(b) and also Eq.~(\ref{dpB}). 

Bohmian correlation measure defined by $P_B=B(\mathbf{a},\mathbf{b})$ thus for any setup of polarizators violates the Bell inequality  Eq.~(\ref{chsh}) exactly as the quantum counterpart $P$ does. It is well known that Bohmian and standard approach of quantum mechanics are isomorphic regarding the prediction of experimental outcomes therefore  the equality of spin-spin correlations in both approaches apparently is not a surprise. However, 
one should remark,  that the inequality Eq.~(\ref{chsh}) is derived under the assumption of discreteness of the outcome of polarization measurements, while in the present Bohmian analysis the spin of rotors and the corresponding projections can take any value. More importantly, $P_B$ corresponds to instant correlations of an undisturbed rotor pair, {\it prior} to accomplished measurement, not to the probabilities for particular outcomes of discrete values obtained after the measurement of spin (such probabilities are trivially identical in all approaches). 
Therefore, although the correlators of the continuously distributed Bohmian angular momenta are identical to the quantum counterparts, those momenta are not directly related to the results of quantum measurements of spin projections of individual qubits, but
during the process of the polarization measurement the entangled rotors adiabatically develop the measured discrete spin values
from pre-measurement distributions of continuous angular momenta.
This result resembles the findings of an alternative analysis of Stern-Gerlach experiments or Einstein-Podolsky-Rosen spin correlations where de Broglie-Bohm causal predictions of experimental outcomes  are identical to standard quantum mechanics results  \cite{stern,dewdney}.

\section{Summary}

We concentrated on time independent properties of a two spin-$\frac{1}{2}$  particle system, as viewed in the Bohmian space of hidden variables.  The state Eq.~(\ref{psi2}) represents one of the simplest many-particle quantum systems exhibiting quantum entanglement and although
it is determined by two parameters only,  the angular momenta
in the Bohmian space display an infinite range of detail which we endeavoured to present by means of various probability distributions and the corresponding average values for the Bohmian ensemble.  The main motivation  was the analysis of entanglement aspects of a qubit pair, in particular 
the exploration of the background of the angular momenta dynamics leading to specific unison motion for a perfectly entangles qubit pair \cite{epl11} which is also the source of the violation of Bell's inequalities. The main properties of two entangled qubits and the results presented in the paper can be summarized as follows: 

(i) Two entangled qubits represented by the quantum state Eq.~(\ref{psi2}) exhibit a distribution of angular momenta with projections forming in the $xy$-plane an average angle $\langle \phi \rangle=\varphi$.

(ii) The ensemble average cosine of this angle is to an excellent  approximation given by $\langle \cos(\phi -\langle \phi\rangle)\rangle\approx\sin \vartheta$. This means that for perfectly entangled states, $\vartheta=\frac{\pi}{2}$, all pairs of momenta in the ensemble form an equal angle $\phi=\varphi$. In general, a higher degree of entanglement is directly related to a pronounced unison motion of angular momenta, {\it i.e.}, with suppressed relative angle fluctuations in agreement with the results of a similar treatment in the framework of standard quantum mechanics \cite{ramsakQM}.

(iii) In addition to the geometrical picture based on relative angles of the angular momenta projections the entanglement of formation can quantitatively be expressed by appropriate Bohmian probability distributions and the corresponding Shannon entropy.

(iv) The virial theorem for a general system of two  rotors is confirmed numerically: the ensemble averaged quantum potential represents one half of the average kinetic energy of the rotors. A general consequence is a known distinction of Bohmian pre-measurement values from after-measurement standard quantum mechanics results. Specifically, spin-spin correlation tensors differ from customary quantum mechanics counterparts by a factor $\sim2/3$.   

(v) For fully entangled qubit pairs the Bohmian spin-spin  probability distributions exhibit power-law tails and are for particular cases given analytically.

(vi) It is shown that the Bohmian analogue of Bell's inequalities, expressed in terms of Bohmian spin-spin correlators, is for fully entangled states identical to the standard quantum mechanics counterpart, which demonstrates the non-local nature of hidden variables in the present approach. The source of non-locality lies in the quantum potential which generates an instant coupling between the angular momenta of entangled qubit pairs. In fully entangled states the angular momenta precess in a particular manner forming a constant relative azimuthal angle $\varphi$ which is the origin of a specific form of both,  the standard quantum mechanical and the Bohmian correlators,
\[
(a_x b_x+a_y b_y)\cos\varphi
+(a_y b_x-a_x b_y)\sin\varphi- 
a_z b_z,
\]
leading to  the violation of the Bell inequalities. It should be noted that the equality of $C(\mathbf{a},\mathbf{b})$ and $B(\mathbf{a},\mathbf{b})$ guarantees
equivalence of Bohmian and standard quantum mechanics regarding the (non)violation of all inequalities which can quantum mechanically be expressed  in terms of the elements of spin-spin correlator tensor normalized as in Eq.~(\ref{cb}), including recently introduced tests for local realistic models \cite{bloblacher07}.

The author thanks T. Rejec, I. Sega, J. H. Jefferson, and T. Huljev {\v C}ade{\v z} for discussions and  he  acknowledges the support from the Slovenian Research Agency under Contract No. P1-0044.


\begin{thebibliography}{10}
\bibitem{komentar1}{Strictly speaking, in the framework of the standard quantum mechanics in most cases the counterparts of the Bohmian probability distributions do not exist and the comparison of both approaches is limited to the discussion of various average values only.}
\end{thebibliography}
\end{document}